\documentclass[twocolumn,showpacs,amsmath,amssymb]{revtex4}

\usepackage{graphicx}
\usepackage{dcolumn}
\usepackage{bm}
\newcommand{\etal}{\emph{et al.}}
\newcommand{\be}{\begin{equation}}
\newcommand{\ee}{\end{equation}}
\newcommand{\bfig}{\begin{figure}}
\newcommand{\efig}{\end{figure}}
\newcommand{\incl}{\includegraphics}

\begin{document}      

\title{Depairing field, onset temperature and the nature of the transition in cuprates
} 
\author{Lu Li$^1$, Yayu Wang$^1$, J. G. Checkelsky$^1$, M. J. Naughton$^2$,\\
Seiki Komiya$^3$, Shimpei Ono$^3$, Yoichi Ando$^3$ and N. P. Ong$^1$\footnote{Proceedings 
M$^2$S-HTSC-VIII, Dresden 2006, Physica, \emph{in press}.}
}
\affiliation{
$^1$Department of Physics, Princeton University, Princeton, NJ 08544\\
$^2$Department of Physics, Boston College, Chestnut Hill, Massachusetts 02467, U.S.A\\
$^3$Central Research Institute of Electric Power Industry, Komae, Tokyo 201-8511, Japan
}

\date{\today}      
\begin{abstract} The depairing (upper critical) field $H_{c2}$ in hole-doped cuprates has been inferred from 
magnetization curves $M$-$H$ measured by torque magnetometry in fields $H$ up to 45 T.  We discuss
the implications of the results for the pair binding energy, the Nernst onset temperature, fluctuations and the 
nature of the Meissner transition at $T_c$.
\end{abstract}

\maketitle                   
In hole-doped cuprates, the depairing field at which the pair condensate is destroyed 
(or ``upper critical field" $H_{c2}$) has been notoriously difficult to measure. 
Recently, progress has been achieved using the vortex-Nernst 
effect~\cite{Wang01,WangSci,Wang06} and high-field torque magnetometry~\cite{Li05,Wang05}.

High-field measurements of $M$ are technically difficult because the large Ginzburg-Landau 
parameter, small crystal volumes (0.1-0.3 mm$^3$) and high field scales of $H_{c2}$ (50-200 T) 
all result in a very small sample moment.  Fortunately, torque magnetometry is well-suited for this purpose~\cite{Bergemann}.  
The crystal is glued to the end of a soft cantilever with its axis $\bf c$ at a small angle to $\bf H$. 
The observed magnetization $M_{eff} = \Delta\chi H_z + M(T,H_z)$,
where $M(T,H_z)<0$ is the magnetization produced by supercurrents ($\bf\hat{z}||c$ and $T$
is the temperature).  The dominant contribution to the paramagnetic background
$\Delta\chi$ comes from the strongly anisotropic orbital (van Vleck) term 
$\chi_{i}^{orb}$.  Its weak $T$-linear behavior allows the diamagnetic term $M$
to be extracted with high resolution~\cite{Wang05}.  
Here we discuss some of our torque measurements on
$\rm La_{2-x}Sr_xCuO_4$ (LSCO), $\rm Bi_2Sr_{2-y}La_yCuO_6$ (Bi 2201) 
and $\rm Bi_2Sr_2CaCu_2O_8$ (Bi 2212) from the underdoped (UD) to overdoped (OD)
regimes.  

\bfig[h]            
\incl[width=8.5cm]{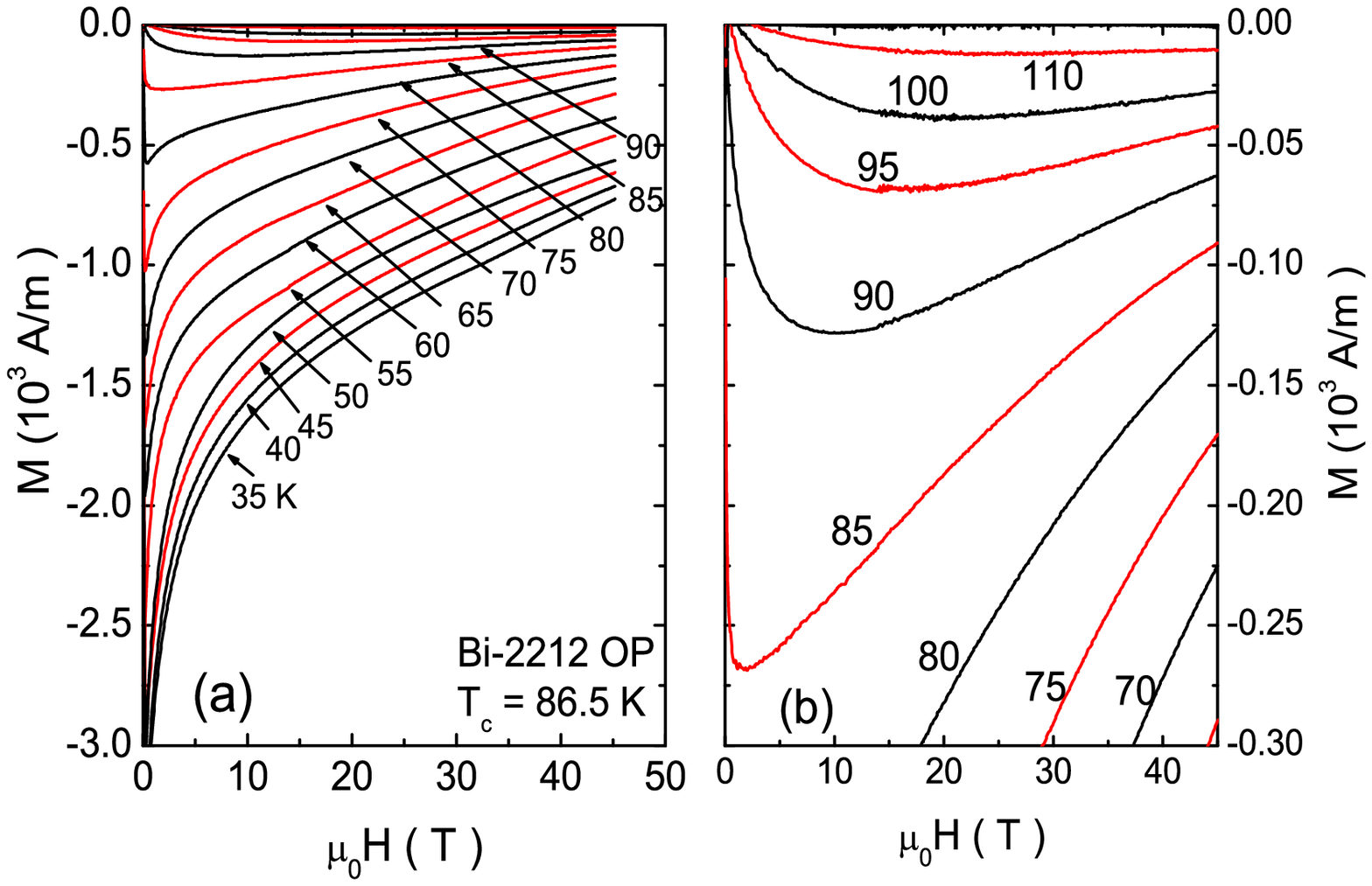}
\caption{\label{MH2212} Magnetization curves $M$ vs. $H$ in OP Bi 2212 at 
$T$ = 35-90 K (Panel a) and at $T$ = 75-110 K (b).  At low $T$, $|M|\sim \log H$ initially,
but goes to zero at $H_{c2}$ = 150-200 T.  Notably, $H_{c2}(T)$ shows no
tendency towards zero as $T\rightarrow T_c^{-}$.  Above $T_c$ (86.5 K),
$|M|$ remains quite large in fields up to and above 45 T.
}
\efig

Figure \ref{MH2212} shows the $M$-$H$ curves in optimally-doped 
(OP) Bi 2212 ($T_c\sim$ 86.5 K) at temperatures 35 to 90 K (left 
panel), and from 80 to 110 K (right panel).  The curves are all fully reversible (pinning 
effects appear only below 35 K at low fields $<$ 1 T).
At the lowest $T$, the curve of $M$ vs. $H$ is very similar
to that in a low-$T_c$ type II superconductor.  Above $H_{c1}$,
$|M|$ decreases as $\log H$ over a very broad field range.  A notable 
feature is the very high $H_{c2}$, which we estimate to be 150-200 T by extrapolation.  
(These values are much larger than inferred from measurements of the resistivity $\rho$ vs. $H$.
The ``knee'' feature in $\rho$ usually used to fix ``$H_{c2}$" actually occurs just above
the vortex-solid melting field $H_m\ll H_{c2}$.)

As $T$ nears $T_c$, a major difference from BCS superconductors emerges.  
There, $H_{c2}(T)$ decreases to zero linearly, viz.
$H_{c2}(T)\sim t$ with $t = 1-T/T_c$.  Accordingly, the high-field 
limit of $M$ in Fig. \ref{MH2212}b should decrease and reach zero at $T_c$.  
Instead, we find that it remains high above our maximum field 
(45 T), even when $T$ exceeds $T_c$ (right panel).  The diamagnetic signal remains 
quite large at 45 T up to 110 K.

\bfig[h]            
\incl[width=8.5cm]{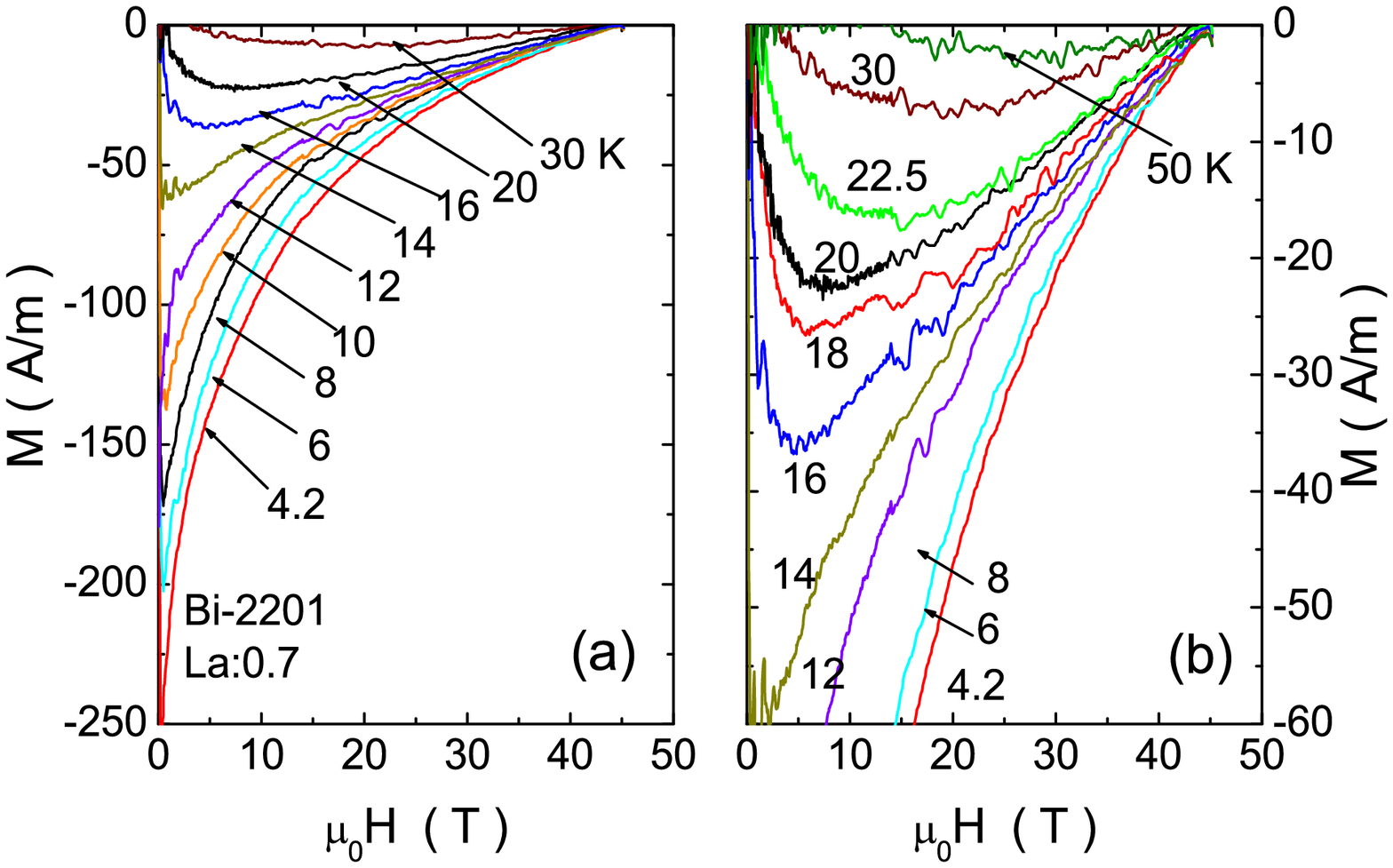}
\caption{\label{MH2201UD} Magnetization curves $M$ vs. $H$ in UD Bi 2201 shown
for $T$ = 4.2-30 K (Panel a), and in expanded scale (Panel b).  At each $T$, the field at 
which $|M|\rightarrow 0$ is taken to be $H_{c2}(T)$.  The convergence of all curves
implies that $H_{c2(T)}$ is independent of $T$ to our resolution.  Above $T_c\sim$ 14 K,
$M$ remains sizeable and strongly $H$ dependent.
}
\efig

This key feature  -- seen in all the hole-doped
cuprates studied -- is most apparent in single-layer UD Bi 2201, where 
complete suppression of diamagnetism is attainable below 45 T.  
Figure \ref{MH2201UD} shows the $M$-$H$ curves in a crystal 
with $T_c\sim$ 14 K.  Hysteretic behavior is not observed down to 4 K.
In comparison with OP Bi 2212, the magnitude $|M|$ in weak $H$ and low $T$
is quite a bit smaller (250 A/m compared with 4000 A/m), but it displays the same $\log H$ dependence
in $H<$ 20 T.  At high fields, $M$ approaches zero at the field $H_{c2}(0)\sim$ 43 T.  In Panel b,
we show that $H_{c2}(T)$ remains nominally $T$-independent even above $T_c$.   
In the interval around $T_c$, the $M$-$H$ curves display 
the same pattern as shown for OP Bi 2212.  Significant diamagnetism remains
at $T$ up to 30 K.

\bfig            
\incl[width=7cm]{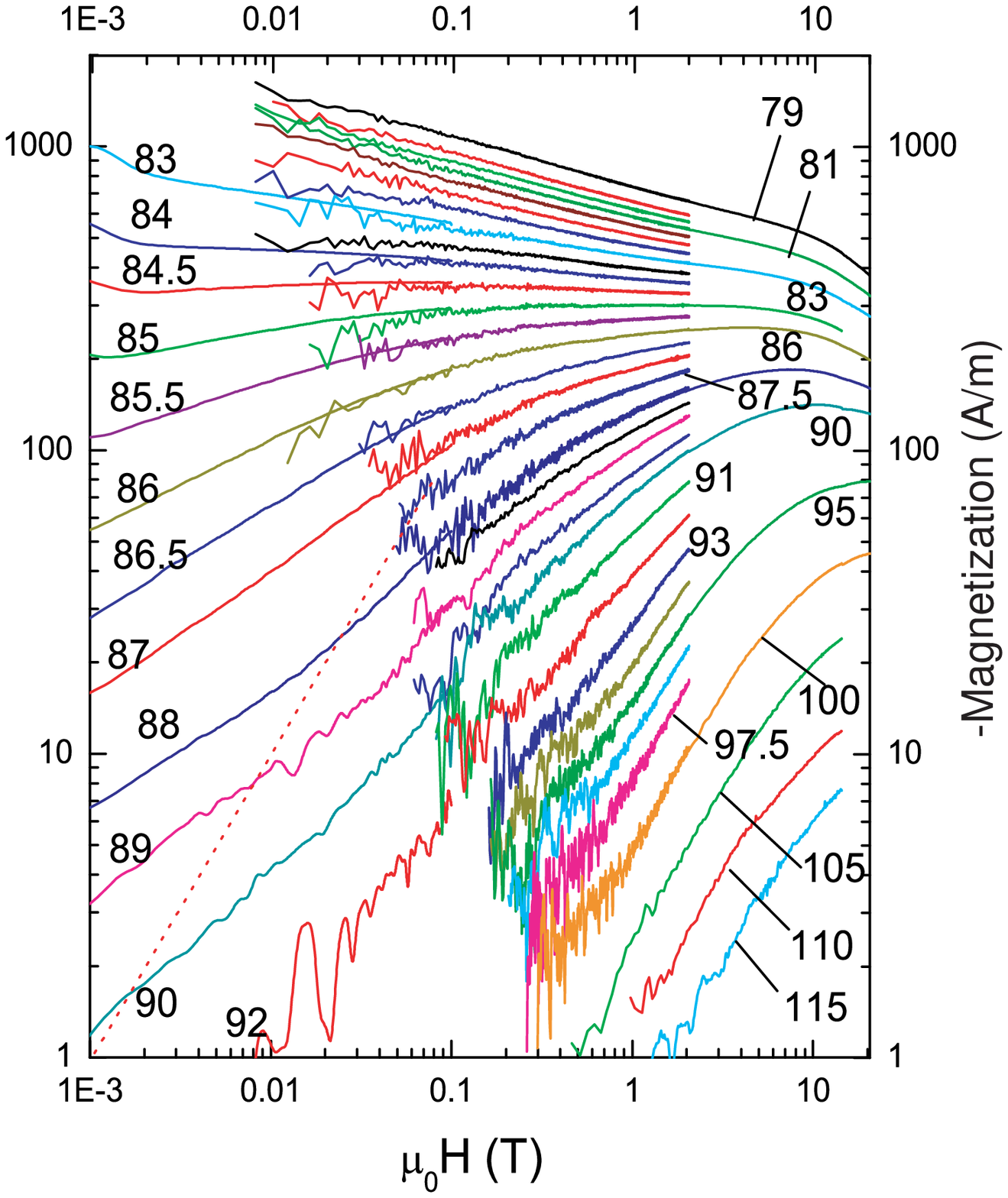}
\caption{\label{loglog} 
The field dependence of $M(T,H)$ in OP Bi 2212 
from $H$ = 10 Oe to 20 T at $T$ = 79-115 K.  The log-log plot shows that, 
as $H\rightarrow 0$, $M(H)$ follows the fractional power-law $M\sim H^{1/\delta}$.
The plot combines SQUID results (10 to 1000 Oe) and toque magnetometry 
results (500 Oe to 2 T).  Torque results up to 20 T are also shown at selected $T$.  
Linear response $M\sim H$ at 87 K would appear as the dashed line
(from Ref. \cite{Li05}).
}
\efig

The diamagnetic signal $M$ above $T_c$ is robust to fields of 45 T
and considerably larger in magnitude than ``fluctuating diamagnetism" in low-$T_c$
superconductors~\cite{Gollub}.  We discuss why the fluctuations are distinctly non-Gaussian.
In BCS superconductors, a key feature is the linear 
decrease to zero of the $T$-dependent critical field $H_{c2}(T)$ as $T\rightarrow T_c^{-}$.
This feature dictates the behavior of fluctuations above $T_c$ in 
Gaussian GL treatments.  In particular, $t$ enters in $M(t,h)$ as the ratio 
$|t|/h$.  Consequently, above $T_c$, the field dependence of $M$ is dictated by 
the field scale $H_{c2}(0)t$, which is the ``mirror image" of $H_{c2}(T)$,
vanishing linearly in $|t|$ as $T\rightarrow T_c$ from above.  Accordingly,
$M$ measured in low-$T_c$ superconductors are nicely scaled when plotted in terms
of the ``Prange" variable $x = [\frac{dH_{c2}}{dT}]_c(T-T_c)/H\sim 1.4|t|/h$~\cite{Gollub}.

As noted above, $H_{c2}(T)$ does not vanish linearly in $|t|$ in hole-doped 
cuprates~\cite{Wang06,Li05,Wang05}.  This invalidates the Gaussian approach which
depends on series expansion in terms of the order parameter and its derivative.  Above $T_c$, the $M$-$H$
curves in Bi 2212 are also qualitatively different from low-$T_c$ superconductors.  Instead of
linear response, the $M$-$H$ curves are strongly nonlinear even in low $H$.  Figure \ref{loglog} shows 
in log-log scale the variation of $M$ over a broad range of $H$ (10 G to 30 T) at $T$ = 79-115 K.  As
emphasized in Fig. \ref{delta}a, the $M$-$H$ curves display strong curvature at
temperatures near $T_c$ = 87 K.  In weak $H$,
$M(H)$ fits well to the power law 
\be 
|M|\sim H^{1/\delta},
\label{MH}
\ee
with a strongly $T$ dependent exponent $\delta(T)$.  Between $T_c$ and 105 K, $\delta(T)$ 
decreases from $\sim 10$ to 1 (Fig. \ref{delta}b).  
The vanishing of $\delta(T)$ defines the temperature $T_s$ slightly below $T_c$ at which 
$M$ is \emph{independent} of $H$ up to a few T (this feature -- dubbed the separatrix~\cite{Li05} --
has been known for a long time).  

These anomalous magnetization patterns are incompatible with Gaussian fluctuations, but
consistent with the phase-disordering scenario~\cite{Kivelson} in 
which, above $T_c$, the condensate amplitude is large, but phase rigidity and 
long-range phase coherence are lost.  Near $T_c$, the $M$-$H$ curves are strikingly 
similar to those calculated for a 2D superconductor near its Kosterlitz Thouless 
(KT) transition~\cite{Vadim}.  As discussed later, the appropriate comparison is
with the 3DXY model with very large anisotropy.

\bfig            
\incl[width=8.5cm]{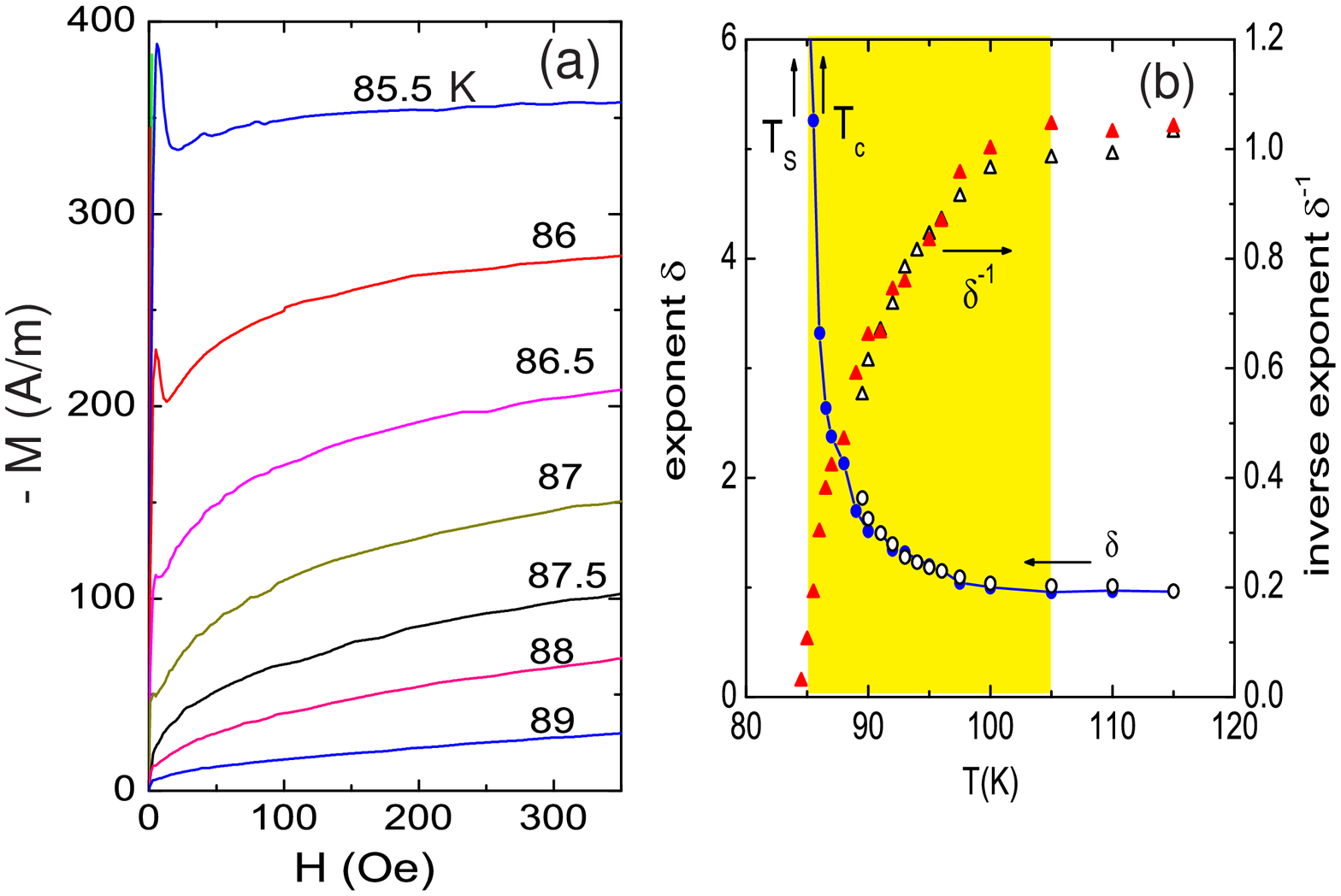}
\caption{\label{delta} (Panel a) The weak-field $M$-$H$ curves near $T_c$
(= 87 K) in OP Bi 2212.  The power-law variation with fractional
exponent is evident in all curves shown.  Below $T_c$, full flux expulsion occurs
for $H<H_{c1}$ (visible as a spike).  The curve at 85.5 K is very close to the
separatrix temperature $T_s$.  Panel (b) displays
the $T$ dependence of the exponent $\delta(T)$ (circles)
for 2 crystals of OP Bi 2212.  The reciprocal 
$\delta(T)^{-1}$ is plotted as solid and open triangles for the 2 samples.
As $T\rightarrow T_s^{+}$, $\delta(T)^{-1}$ decreases smoothly to 0.  In the shaded region 
where $\delta > 1$ (from $T_s$ to 105 K), linear magnetic response is absent even 
at 10 Oe (adapted from Ref. \cite{Li05}).
}
\efig

The $M$-$H$ curves in Figs. \ref{MH2212}-\ref{loglog} together imply the following
physical picture (see Ref. \cite{Li05}).  On cooling from 300 K, 
the system first crosses the pseudogap temperature $T^*$.
The pseudogap affects primarily the spin degrees of freedom, especially the 
relaxation rate $1/T_1T$ in NMR and the bulk susceptibility.  
Evidence for Cooper charge pairing appears only at $T_{onset}$ = 0.5-0.7 $T^*$.  
Below $T_{onset}$ (the ``Nernst" region), both the Nernst signal and diamagnetism increase
steeply to merge smoothly with the corresponding signals below $T_c$.  
Within each CuO$_2$ layer, the pair condensate is robust with a 
very large pair-binding energy.  However, because of thermal generation of 
mobile 2D vortices, phase coherence is confined to a length 
scale given by the phase correlation length $\xi_{\phi}$ ~\cite{Li05}.  The ``hot
2D vortex liquid", nonetheless, displays a fairly large diamagnetic response.

It is instructive to contrast cuprates from a percolative system (e.g. granular Al)
in which superconducting islands are gradually phase coupled by the proximity effect
as $T$ decreases.  In high-quality crystals of the cuprates, the zero-field transition is invariably
very sharp.  At $T_c$, full flux expulsion appears~\cite{Li05}.  The resistive
transition is also abrupt, in contrast to the long tail seen in granular Al.

Two features seem to be crucial.  The first is the pre-emption of the 2D KT
transition in individual layers by the 3D transition caused by interlayer coupling,
as occurs in layered magnets~\cite{Hirakawa}.
Below $T_{onset}\sim$ 130 K, the in-plane $\xi_{\phi}$ (inferred from
the susceptibility $\chi = M/H$) grows as in the KT transition~\cite{Li05}.  Below 105 K, however,
$M$ becomes increasingly non-linear (Fig. \ref{delta}a).  The 
increase in $\delta(T)$ reflects rapid upward renormalization of the interlayer coupling
strength.  In the interval between $T_c$ and 105 K, the fractional power-law implies 
that $\chi$ can approach -1 in the limit $H\rightarrow 0$.  
However, this London rigidity is fragile and easily suppressed by field.  At $T_c$ ($\sim$2 K above
$T_s$), the Meissner state appears~\cite{Li05}.  As apparent in Fig. \ref{delta}a, 
full flux expulsion occurs below the lower critical field $H_{c1}(T)$, seen as 
sharp spikes in Fig. \ref{delta}a.  Significantly, the 3D Meissner state is 
observed at fields $H<H_{c1}$, yet at higher $H$ ($>$ a few T), 
the $M$-$H$ profiles revert to the 2D pattern seen high above $T_c$.  
This field-induced crossover from 3D to 2D behavior -- with its intrinsic
non-monotonicity -- is very different from low-$T_c$ superconductors.

The second feature is the termination of the melting curve $H_m(T)$ at $T_c$.  
Far from being accidental, we believe this is intrinsic to the nature of the
transition.  Within the vortex-solid state, spontaneously 
created vortices are ineffectual in destroying phase coherence because
they are not able to diffuse.  Hence, $T_c$ cannot lie below the high-$T$ 
termination of $H_m(T)$.  On the other hand, $T_c$ cannot lie above
the termination point.  This would correspond to a strictly 2D transition
that is incompatible with the observed full Meissner effect.  
These 2 features distinguish the cuprate transition from that in
granular Al.

\bfig[h]            
\incl[width=7cm]{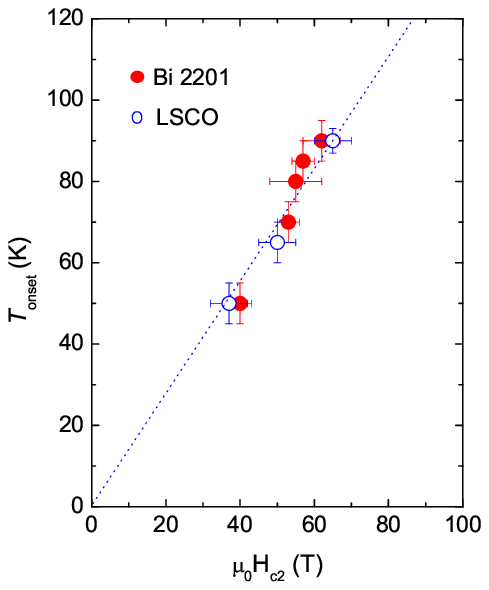}
\caption{\label{Tonset} 
Plot of $T_{onset}$ vs. $H_{c2}(0)$ in the single-layer cuprates Bi 2201 
and LSCO.  Both quantities are inferred from the $M$ vs. $H$ curves measured
by high-field torque magnetometry.  The broken line is Eq. \ref{gmu}
with $g\simeq$ 2.1.
}
\efig

Lastly, we discuss an interesting relation between $H_{c2}$ and $T_{onset}$.
Measurements on several Bi 2201 and
LSCO crystals from UD to OD regime reveal that $H_{c2}(0)$ and 
$T_{onset}$ (measured from both the Nernst signal and magnetization) scale together
as shown in Fig. \ref{Tonset}.  Within the experimental uncertainties, $T_{onset}$
is linear in $H_{c2}(0)$.  Expressing $H_{c2}(0)$ as a Zeeman energy, viz.
\be
k_BT_{onset} = g\mu_BH_{c2}(0),
\label{gmu}
\ee
we find that the g-factor $g\simeq$ 2.1 ($\mu_B$ is the Bohr magneton).  The data
in Fig. \ref{Tonset} are restricted to Bi 2201 and LSCO.  As noted in Fig. \ref{MH2212},
$H_{c2}(0)$ in OP Bi 2212 is much larger given its $T_{onset}$ ($\sim$130 K).
Figure \ref{Tonset} ties together the energy scale implied by $T_{onset}$
and the pair binding energy at low $T$ for Bi 2201 and LSCO.  
The linear fit with $g\sim$ 2.1 suggests that the Pauli limit
may be relevant to the high-field depairing process.  The implication of this relationship
is currently being explored.

We acknowledge discussions with P. W. Anderson, S. A. Kivelson, Z. Tesanovic, V. Oganesyan, 
S. Sondhi, D. A. Huse and Z. Y. Weng.  Measurements were performed at the National High Magnetic Field 
Laboratory, Tallahassee, which is supported by the U.S. National Science 
Foundation (NSF DMR-0084173), the State of Florida and DOE.  This research is supported by 
NSF Grant DMR 0213706.

\end{document}